\def \SAIT #1 #2 {{\em Mem.\ Soc.\ Astron.\ It.\/} {\bf #1}, #2}
\def \MESS #1 #2 {{\em The Messenger\/} {\bf #1}, #2}
\def \ASTRNACH #1 #2 {{\em Astron. Nach.\/} {\bf #1}, #2}
\def \AAP #1 #2 {{\em Astron. Astrophys.\/} {\bf #1}, #2}
\def \AAL #1 #2 {{\em Astron. Astrophys. Lett.\/} {\bf #1}, L#2}
\def \AAR #1 #2 {{\em Astron. Astrophys. Rev.\/} {\bf #1}, #2}
\def \AAS #1 #2 {{\em Astron. Astrophys. Suppl. Ser.\/} {\bf #1}, #2}
\def \AJ #1 #2 {{\em Astron. J.\/} {\bf #1}, #2}
\def \ANNREV #1 #2 {{\em Ann. Rev. Astron. Astrophys.\/} {\bf #1}, #2}
\def \APJ #1 #2 {{\em Astrophys. J.\/} {\bf #1}, #2}
\def \APJL #1 #2 {{\em Astrophys. J. Lett.\/} {\bf #1}, L#2}
\def \APJS #1 #2 {{\em Astrophys. J. Suppl.\/} {\bf #1}, #2}
\def \APSS #1 #2 {{\em Astrophys. Space Sci.\/} {\bf #1}, #2}
\def \ASR #1 #2 {{\em Adv. Space Res.\/} {\bf #1}, #2}
\def \BAIC #1 #2 {{\em Bull. Astron. Inst. Czechosl.\/} {\bf #1}, #2}
\def \JSQRT #1 #2 {{\em J. Quant. Spectrosc. Radiat. Transfer\/} {\bf #1}, #2}
\def \MN #1 #2 {{\em Mon. Not. R. Astr. Soc.\/} {\bf #1}, #2}
\def \MEM #1 #2 {{\em Mem. R. Astr. Soc.\/} {\bf #1}, #2}
\def \PLR #1 #2 {{\em Phys. Lett. Rev.\/} {\bf #1}, #2}
\def \PASJ #1 #2 {{\em Publ. Astron. Soc. Japan\/} {\bf #1}, #2}
\def \PASP #1 #2 {{\em Publ. Astr. Soc. Pacific\/} {\bf #1}, #2}
\def \NAT #1 #2 {{\em Nature\/} {\bf #1}, #2}
\def\ga{\mathrel{\mathchoice {\vcenter{\offinterlineskip\halign{\hfil
$\displaystyle##$\hfil\cr>\cr\sim\cr}}}
{\vcenter{\offinterlineskip\halign{\hfil$\textstyle##$\hfil\cr  
>\cr\sim\cr}}}
{\vcenter{\offinterlineskip\halign{\hfil$\scriptstyle##$\hfil\cr
>\cr\sim\cr}}}  
{\vcenter{\offinterlineskip\halign{\hfil$\scriptscriptstyle##$\hfil\cr
>\cr\sim\cr}}}}}

\documentstyle{memsait}
\input epsf.sty
\begin{opening}
\title{AGB EVOLUTION WITH OVERSHOOT: \\HOT BOTTOM BURNING AND DREDGE UP}
\author{T.\,BL\"OCKER$^1$,
F.\,HERWIG$^2$, AND
T.\,DRIEBE$^1$}
\institute{
$^1$Max--Planck--Institut f\"ur Radioastronomie, Bonn, Germany\\
$^2$Institut f\"ur Physik, Universit\"at Potsdam, Germany 
}
\date{} 
\end{opening}

\begin{document}

\oddpagefooter{}{}{} 
\evenpagefooter{}{}{} 
\ 
\bigskip

\begin{abstract}
We calculated models of massive AGB stars with a self-consistent coupling of
time-dependent mixing and nuclear burning for 30 isotopes and 74 reactions.
Overshoot with an exponentially declining velocity field was considered
and applied during all stages of evolution and in all convective regions.
Very efficient $3^{\rm rd}$ dredge-up was found even overcompensating
the growth of the hydrogen-exhausted core after a few thermal pulses.
Hot bottom burning occurs for $M\ge 4\,$M$_{\odot}$ within the sequences with
overshoot. Carbon star formation in these more massive AGB stars is delayed
or even prevented by hot bottom burning despite the very efficient dredge-up.
With the simultaneous treatment of mixing and burning the 
formation of Li-rich AGB stars due to the Cameron-Fowler mechanism was
followed. For  a $6\,$M$_{\odot}$ model
the maximum Li abundance was found to be
$\epsilon(^7{\rm Li}) \approx 4.4$. 
\end{abstract}

\section{Introduction}
In recent stellar evolution calculations Herwig et al.\ (1997)
have considered diffusive overshoot for all convective boundaries which
leads to a considerable change in the models.
This method provides for Asymptotic Giant Branch (AGB) 
stars a sufficient amount of dredge-up
to form low-mass carbon stars as required from observations.
Additionally it leads to the formation of $^{13}$C as a neutron source to
drive the s-process in these stars. More details are given in
Herwig et al.\ (1999).

However the consequences for
massive AGB stars, which suffer from hot bottom burning, remain to be 
investigated in detail.
Hot bottom burning, i.e.\ the penetration of the convective envelope into the
hydrogen burning shell, provides lithium-rich AGB stars
and may delay or prevent the  carbon star stage 
by turning $^{12}$C into $^{13}$C and  $^{14}$N. 
We have extended our computations of AGB stars
with diffusive overshooting to hot bottom burning models
including a self-consistent coupling of nucleosynthesis and mixing.
\section{Computational details}
We used the evolutionary code of Bl\"ocker (1995).
Nuclear burning was accounted for by a nuclear network covering
30 isotopes and 74 reactions up to carbon burning. We considered the
opacities of Iglesias \& Rogers (1996) supplemented with those of
Alexander \& Ferguson (1994) for the low temperature regime.

Convection was treated within the mixing length theory (B\"ohn-Vitense 1958)
with a mixing-length parameter $\alpha=1.7$. Mixing of chemical elements
was treated by solving a diffusion equation. 
Overshoot was taken into account according to the prescription of
Herwig et al. (1997) which is based
on the hydrodynamical calculations of Freytag et al. (1996).
Freytag et al. (1996) showed that mixing takes place well beyond the classical
Schwarzschild border due to overshooting convective elements 
with an exponentially declining velocity field.
In the overshoot region the corresponding diffusion coefficient is given by
$D_{\rm os}=v_0\cdot H_{\rm p}\cdot\exp\frac{-2z}{f\cdot H_{\rm p}}$
with $v_0$: velocity of the convective elements immediately before the
Schwarzschild border; $z$: distance from the edge of the convective zone;
$f$: the overshoot efficiency parameter. We used an
efficiency parameter of $f=0.016$ as appropriate to match the observed
width of the main sequence. Diffusive overshooting was considered 
in all convective regions during the complete evolution.

Abundance changes have been treated self-consistently by coupling
time-dependent mixing and nuclear burning of all chemical elements,
i.e. by solving
\begin{displaymath}\label{sbm}
\frac{{\rm d}X_i}{{\rm d}t}=\left(\frac{\partial X_i}{\partial
t}\right)_{\rm nuc} +
\frac{\partial}{\partial m}\left[(4\pi r^2 \rho)^2 D\frac{\partial
X_i}{\partial m}\right]_{\rm mix}
\end{displaymath}
The simultaneous treatment of burning and mixing is essential to follow, e.g.\ 
the $^7{\rm Li}$-production in luminous AGB stars via the Cameron-Fowler
mechanism (Cameron \& Fowler 1971, see also Sackmann \& Boothroyd 1992).
The self-consistent solution of time-dependent burning and mixing processes
is, however, associated with  a considerable increase of computing time
due to the size of the problem (30 isotopes, $\sim 2000$ mass shells).
We calculated sequences for 
initial masses of $3 {\rm M}_\odot$ to $ 6\,{\rm M}_\odot$ and
$(X,Y)=(0.70,0.28)$ from the pre-main sequence stage up to the AGB and
through the
thermal pulses. In order to disentangle the influence of mass loss 
from the one of overshoot we refer in this study only to sequences
with small mass-loss rates (Reimers 1975, $\eta=1.0$). Complete sequences
will also consider stronger (and better suited) rates
(cf. Bl\"ocker 1995).
%
%
%

\section{Thermal pulses and third dredge up}
On the upper AGB the helium burning shell becomes recurrently unstable
raising the thermal pulses 
(Schwarzschild \& H\"arm 1965, Weigert 1966). During these
instabilities the luminosity of the He shell increases rapidly for a
short time of 100\,yr to $10^{6}$ to $10^{8}$\,L$_{\odot}$.
The huge amount of energy produced forces the development of a pulse-driven
convection zone which mixes products of He burning, carbon and oxygen,
into the intershell region.
Because the H shell is pushed concomitantly into cooler domains
H burning ceases temporarily allowing the envelope convection to proceed
downwards after the pulse,
to penetrate those intershell regions formerly enriched with
carbon and to mix this material to the surface (3$^{\rm rd}$ dredge up,
see, e.g., Bl\"ocker 1999 for a recent review).
Overshoot leads to an enlargement of the pulse-driven convection zone and
to enhanced mixing of core material from deep layers below the He shell
to the intershell zone
(``intershell dredge-up'') and to a deepening of the envelope convection. 
If the determination of convective boundaries is solely based on the
Schwarzschild criterion as in our case,
dredge up can easily be obtained if some envelope overshoot is present
to overcome the H/He discontinuity.
The total amount of dredge up, however, depends mainly on the strength of
the former intershell dredge-up (cf.\ Herwig et al.\ 1999).
As in the case of low-mass stars intershell dredge-up leads to considerable
changes of the intershell abundances. After intershell dredge-up
the abundances (mass fractions) of He, C, O amount to (40,40,14) instead
to (70,25,2) as in non-overshoot sequences. 

%
\begin{figure}
\begin{center}
\epsfxsize=0.8\textwidth
\epsfbox{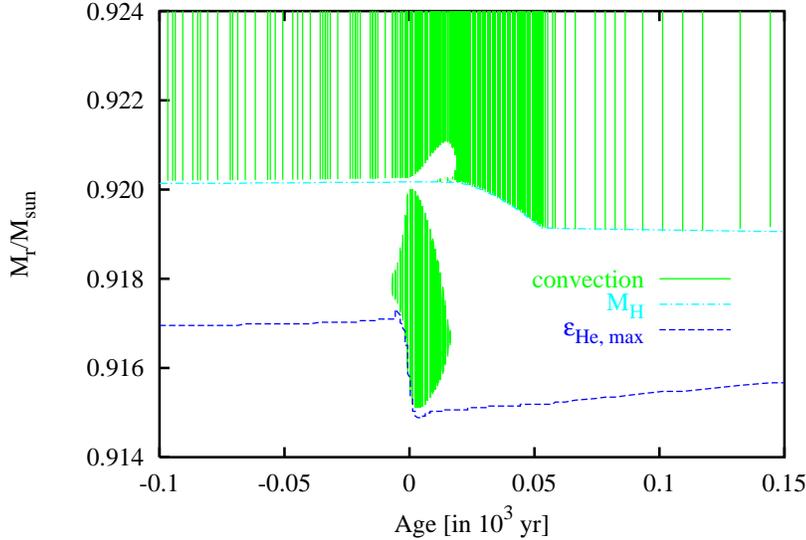}
\end{center}
\vspace*{-8mm}
\caption{Temporal evolution of the extension of convective regions
         for a 6\,M$_{\odot}$ model during the second thermal pulse.
         The hatched regions refer to the bottom of envelope convection
         and to the pulse-driven convective zone of the He shell. The
         dashed-dotted 
         line refers to the core mass $M_{\rm H}$, and the dashed one to
         regions of maximum
         energy production $\varepsilon_{\rm He,max}$ in the He shell.}
\label{Fkiptbl6m}
\end{figure}
\vspace{0.5cm}
\begin{figure}
\epsfxsize=0.9\textwidth
\epsfbox{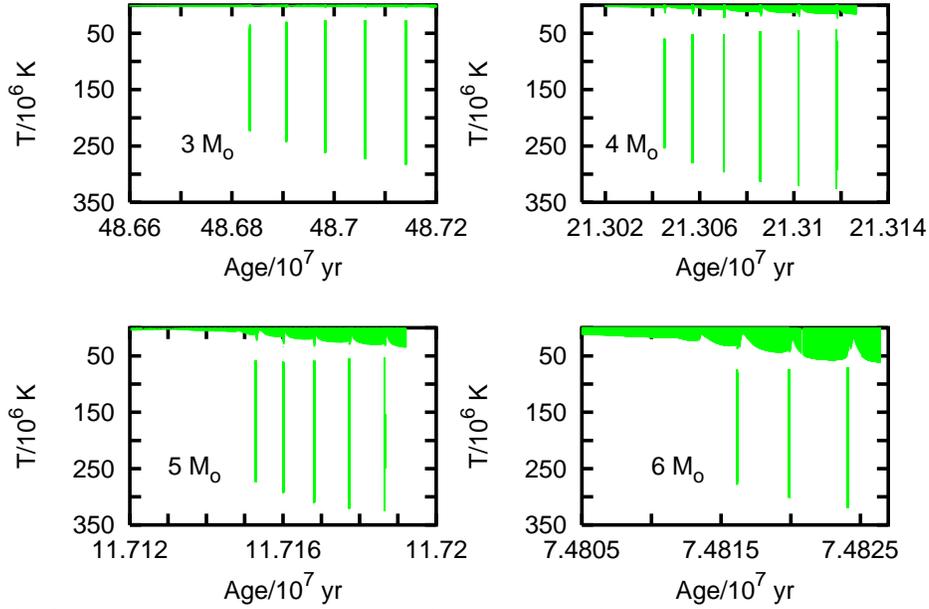}
\caption{Temperature vs. mass for convective regions for different initial
masses.
The hatched regions refer to the bottom of envelope convection
and the ``spikes'' to the pulse-driven convective zone of the He shell.}
\label{Ftpulsmax}
\end{figure}
\begin{figure}[ht]
\vspace{-0.5cm}
\epsfxsize=0.8\textwidth
\epsfbox{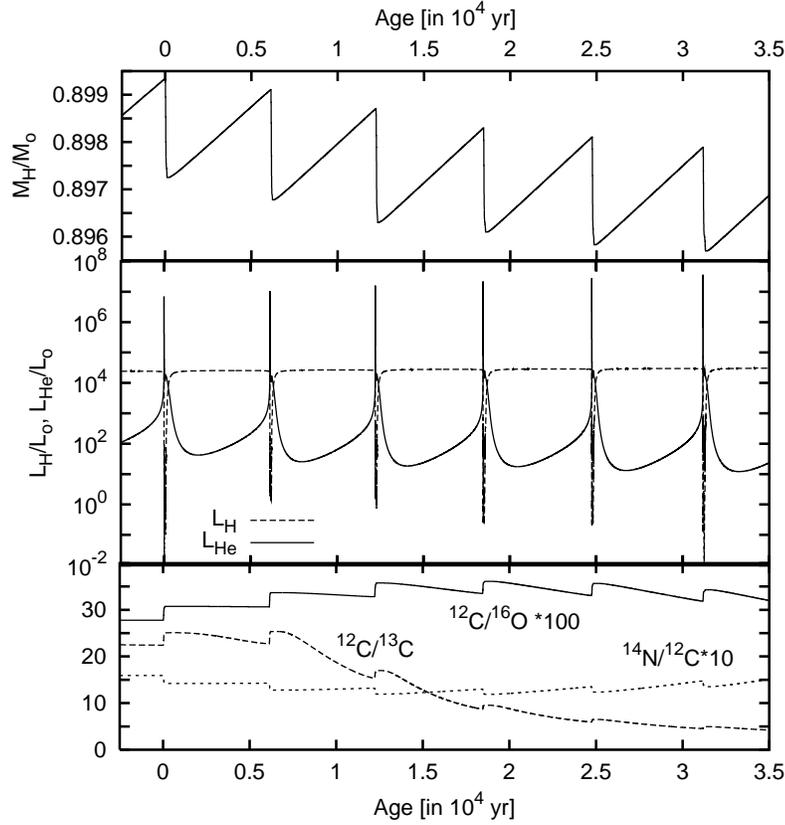}
\caption{Temporal evolution of the core mass  $M_{\rm H}$,
         the shell luminosities $L_{\rm H}$ and $L_{\rm He}$, 
         and surface CNO isotopic ratios (from top to bottom)
         for a 5\,M$_{\odot}$ sequence.}
\label{Fmhydlumi}
\end{figure}

Fig.~\ref{Fkiptbl6m} shows how the lower boundary of the envelope
convection as well as the pulse-driven  convection zone evolve during a pulse
for a 6\,M$_{\odot}$ model. 
The introduction of overshoot and its application to all convective boundaries
provides $3^{\rm rd}$ dredge up even for low-mass
AGB stars (Herwig et al. 1997, 1999)
and we found very efficient dredge up
for higher masses as well. With the dredge-up parameter $\lambda$ being
the ratio of material mixed up during the dredge up
and the material burnt during two
consecutive thermal pulses we find $\lambda> 1$ already during the very first
pulses. Correspondingly, one observes a decrease of
the mass of the hydrogen-exhausted core M$_{\rm H}$ (see Fig.~\ref{Fmhydlumi}).

Due to the overshoot we found high
temperatures at the bottom of the pulse-driven convection zone.
As demonstrated in Fig.~\ref{Ftpulsmax} one obtains
for a 3\,M$_{\odot}$ sequence
temperatures already close to $300 \cdot 10^{6}$\,K, the threshold temperature
of the neutron source $^{22}$Ne($\alpha$,$n$)$^{25}$Mg to operate.
More massive models show even temperatures of  $350 \cdot 10^{6}$\,K and more
after a few pulses. Thus, the $s$-process nucleosynthesis in these
stars will be governed by both the  $^{13}$C($\alpha$,$n$)$^{16}$O
and the  $^{22}$Ne($\alpha$,$n$)$^{25}$Mg neutron source. 
%
%
%
%
\section{Hot Bottom Burning}
\begin{figure}
\epsfxsize=0.75\textwidth
\hspace*{15mm}
\epsfbox{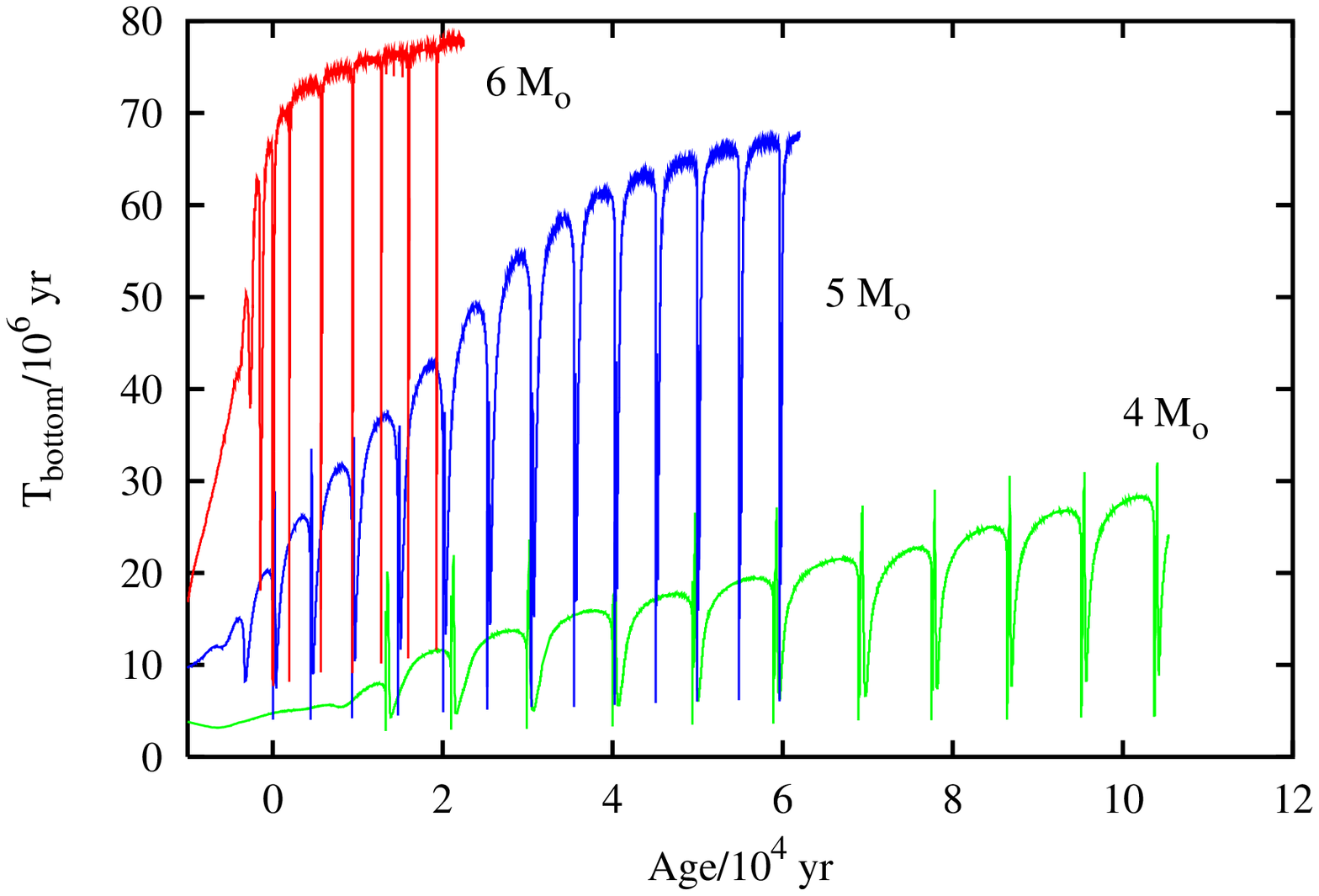}
\caption{Temperature at the base of the convective envelope for models
with initial masses of 4, 5 and 6\,M$_{\odot}$
The strong increase of $T_{\rm bottom}$ indicates the efficiency of HBB.
$t=0$ refers to the $L_{\rm He}$ maximum of the first thermal pulse.
}
\label{Ftkonv456}
\epsfxsize=0.75\textwidth
\hspace*{18mm}
\epsfbox{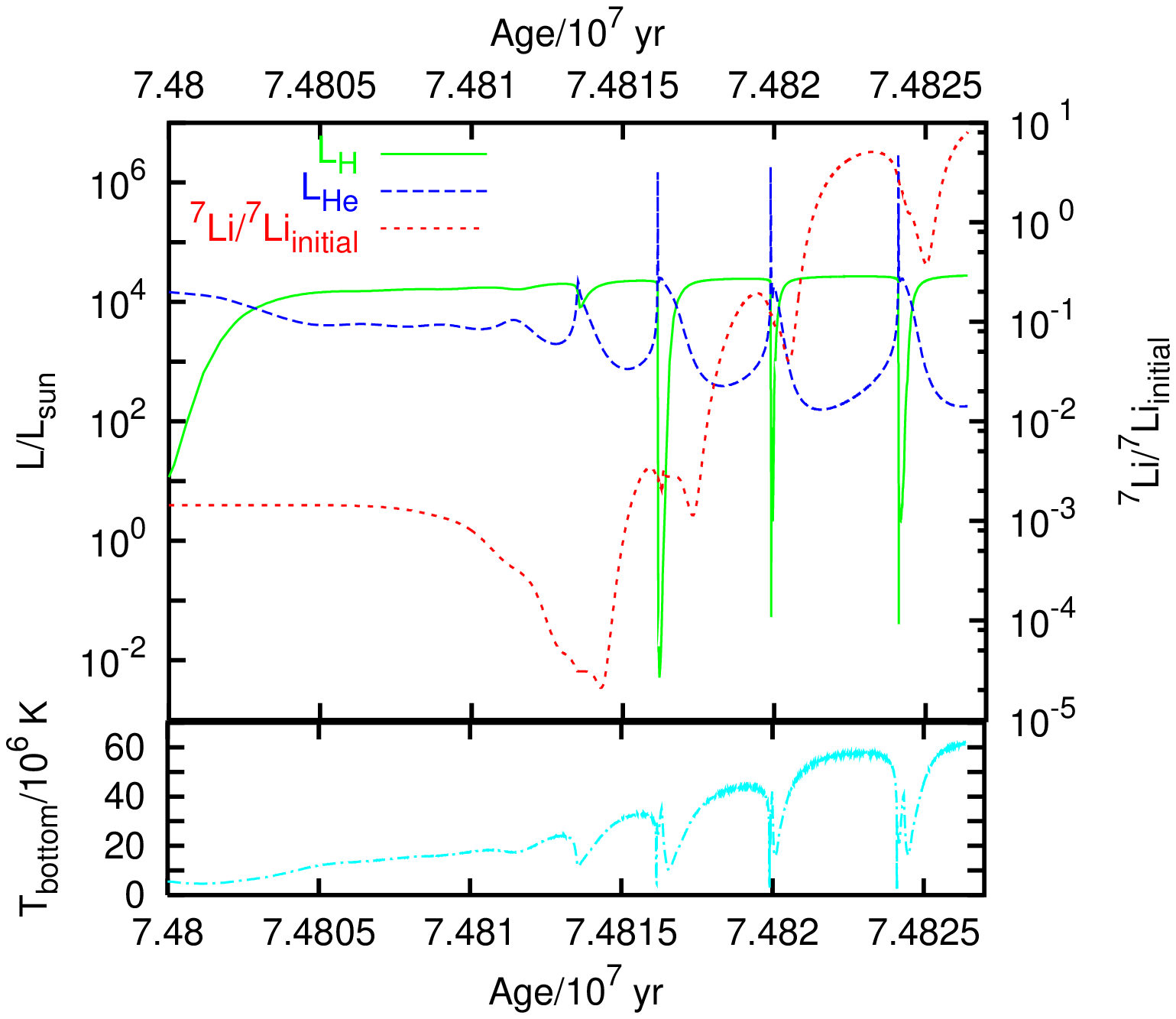}
\caption{
Upper panel: Evolution of the shell luminosities $L_{\rm H}$ and
$L_{\rm He}$,  and of the surface abundance of  $^7{\rm Li}$
with respect to its initial value
($^7{\rm Li}_{\rm initial}=9.35\cdot 10^{-9}$)
during the first thermal pulses of the $M=6 $M$_{\odot}$ sequence.
Lower panel: Corresponding evolution of the temperature at the bottom
of the envelope convection.}
\label{Flilum}
\end{figure}

In more massive AGB stars ($M_{\rm initial} \ga 4$\,M$_{\odot}$) the
convective envelope becomes so extended downwards that it can 
cut into the hydrogen burning shell during the interpulse phase
(hot bottom burning (HBB) or envelope burning; Iben 1975, Scalo et al.\ 1975).
Temperatures in excess of $50 \cdot 10^{6}$\,K (see Fig.~\ref{Ftkonv456})
are reached at the base of
the convective envelope and material burnt there is immediately mixed to the
surface.  HBB models do not obey Paczynski's (1970)
classical core-mass luminosity relation but, instead, evolve rapidly
to very high luminosities (Bl\"ocker \& Sch\"onberner 1991). Due to 
CNO cycling of the envelope $^{12}{\rm C}$ can be transformed into
$^{13}{\rm C}$ and $^{14}{\rm N}$. Consequently, a low
$^{12}$C/$^{13}$C ratio is a typical
signature for HBB which can prevent AGB stars from becoming
carbon stars 
(Iben 1975; Renzini \&  Voli 1981; Boothroyd et al.\ 1993, Frost et al.\ 1998).

The evolution of different CNO isotopic ratios at the surface 
of our 5\,M$_{\odot}$ sequence is shown in the lower panel of
Fig.~\ref{Fmhydlumi}.
The $^{12}{\rm C}/^{13}{\rm C}$ ratio 
reaches its equilibrium value ($\approx 3$) 
after the first pulses and $^{14}{\rm N}/^{12}{\rm C}$ 
starts to increase whereas $^{12}{\rm C}/^{16}{\rm O}$ decreases.
Although dredge-up is very effective leading even to a core-mass decrease,
HBB is so efficient that it prevents or delays the star to become a
carbon star.

The evolution of the surface abundances depends on the
competition between HBB and dredge up.
Both effects  depend on the envelope mass, i.e. on
mass loss. Consequently, the C/O ratio reached at the end of the AGB evolution 
depends crucially on the question which process will shut down first.
The critical minimum enevelope mass for efficient HBB appears to be larger
than the one necessary for dredge up (Frost et al.\ 1998), and it
depends on the remaining dredge-up efficiency if finally carbon stars can be
formed. Our calculations show that dredge up even operates during the
post-AGB stage if overshoot is considered with important conclusions
for the formation of hydrogen-deficient stars.

Another direct consequence of HBB
is the formation of 
lithium-rich stars (Scalo et al.\ 1975) via the
Cameron \& Fowler (1971) mechanism:
The timescale of the $\beta$-decay of $^7{\rm Be}$ which is produced at the
bottom of the envelope convection is comparable to the convective timescale.
Therefore, before $^7{\rm Be}$ is burnt in the lower
part of the envelope it can be mixed up to cooler layers where it decays
to produce $^7{\rm Li}$ which, in turn, is then convected to the surface.
This mechanism is effective for temperatures between
30 and $80 \cdot 10^6\, {\rm K}$ at the bottom of the envelope.
The production of Li-rich AGB stars can only be followed with a simultaneous
treatment of mixing and burning. Otherwise, $^7{\rm Be}$ would be burnt at
the bottom of the envelope before it can be mixed up.

%
Since the luminosity is a unique function of the base temperature and Li
is only produced in a certain temperature range, Li-rich stars  are only
found  between \hbox{$M_{\rm bol}$\,$\approx$\,-6 to -7}
(Sackmann \& Boothroyd 1992)
in good agreement with observations of the Magellanic Clouds
(Smith \& Lambert 1990, Plez et al.\ 1993).

On the other hand, the majority of Li-rich galactic AGB stars
are observed at much lower luminosities, $M_{\rm bol}\approx -3.5$
to $-6$ (Abia \& Isern 1997). They are, however, carbon stars.
%
Since  $^7{\rm Li}$ enrichment requires
$T_{\rm bottom}\ga 30 \cdot 10^6$\,K
whereas $^{12}{\rm C}$ is efficiently destroyed via CNO cycling for
$T_{\rm bottom} \ga 70\cdot 10^6$ K one finds in standard evolution
calculations only a narrow mass, age and
brightness range where AGB stars should be both lithium and carbon rich
indicating that possibly additional mixing processes operate in these stars
(Abia \& Isern 1997).
%

Recently, Ventura et al.\ (1999) calculated AGB models for the LMC with
$M\le 4\, {\rm M}_{\odot}$ considering overshoot in a
similar manner as Herwig et al.\ (1997). They found Li-rich carbon stars
only for $M = 3.5$ and  $3.8\, {\rm M}_\odot$, i.e. for the upper part
of the observed luminosity range. 
%
Our $4\, {\rm M}_\odot$ model ($Z=0.02$) becomes a carbon star
after 10 thermal pulses
due to very
efficient dredge up and can be expected to become lithium rich
during the further course of evolution due to 
the sufficient increase of envelope base-temperatures. 
The $3\, {\rm M}_\odot$ model becomes a carbon star
after a couple of thermal pulses as well but seems to miss the Li-rich stage.
These findings are in line with those of  Ventura et al.\ (1999).
Thus, overshoot alone appears not to be able to explain  
galactic Li-rich carbon for the whole brightness range observed.

Fig.~\ref{Flilum} illustrates the considerable $^7{\rm Li}$ enrichment during
the first thermal pulses for the 6\,M$_{\odot}$ sequence.
Within the first 3 pulses the Li abundance
increases by more than 5 orders of magnitude to reach almost
10 times the main sequence value. 
The maximum lithium abundance is
$\epsilon(^7{\rm Li})=
\log\left[n(^7{\rm Li})/n({\rm H})\right]+12\approx 4.4$ 
in agreement with the results of Sackmann \& Boothroyd (1992).

Mazzitelli et al.\ (1999) calculated lithium production in
AGB stars with application of overshoot to all convective
regions as well.
The overshoot treatment is comparable to the method of
Herwig et al. (1997) but convection is dealt with the
Canuto \& Mazzitelli (1992) prescription.
They obtained a similar evolution of lithium enrichment
and base temperatures but
found for a solar metallicity $6\, {\rm M}_\odot $ model
a complete penetration of the hydrogen burning shell by the
convective envelope resulting in a lack of helium accumulation in the
intershell region and corresponding prevention of thermal pulses.
A corresponding quenching of thermal pulses was not found for the present
$6\, {\rm M}_\odot$ sequence (see Fig.~\ref{Flilum}).

%
%
%
%
\section{Conclusions}
We calculated models of massive AGB stars (solar metallicity)
with a self-consistent coupling of
time-dependent mixing and nuclear burning for 30 isotopes and 74 reactions.
Treating exponential overshoot (Freytag et al. 1996) as in
Herwig et al.\ (1997)  and applying it to all convective regions
we find very efficient $3^{\rm rd}$ dredge-up with
$\lambda > 1$ after a few thermal pulses.
Overshoot leads to intershell dredge-up and provides
considerably changes of the intershell abundances
strongly enriched in carbon and oxygen, i.e. [He,C,O]=[40,40,14] by mass.
These  intershell abundances determine the strength of the
third dredge up, whereas envelope overshoot is mainly required to penetrate
the H/He discontinuity.
Temperatures at the bottom of the flash-driven convection zone of the
helium burning shell are in excess of $300 \cdot 10^{6}$\,K
after a few pulses activating the $^{22}$Ne($\alpha$,n)$^{25}$Mg
reaction for $s$-process nucleosynhesis.

Hot bottom burning occurs for $M\ge 4\,$M$_{\odot}$ within our sequences with
overshoot. Only the  $4\,$M$_{\odot}$ model becomes a carbon star
whereas carbon star formation in more massive AGB stars is delayed
or even prevented by hot bottom burning despite of the very efficient
dredge-up.
With the simultaneous treatment of mixing and burning the
formation of Li-rich AGB stars due to the Cameron-Fowler mechanism was studied.
Already the  $4\,$M$_{\odot}$ model becomes lithium-rich leading to the
formation of a lithium-rich carbon star. 
For  $M\ge 6\,$M$_{\odot}$ the maximum Li abundance was found to be
$\epsilon(^7{\rm Li}) \approx 4.4$.
\vspace*{-1ex}
\acknowledgements
F.H. has been supported by the {\it Deutsche Forschungsgemeinschaft, DFG}
(La\,587/16).

\def \MSSL #1 #2 {{\em Mem. Soc. Sci. Liege\/}  {\bf #1}, #2}
\def \ZAP #1 #2  {{\em Zeitschrift f\"ur Astrophysik\/}  {\bf #1}, #2}
\def \ACA #1 #2  {{\em Acta Astron.\/}  {\bf #1}, #2}

\end{document}